# LASER ABLATION SYNTHESIS OF ZINC OXIDE CLUSTERS:

# A NEW FAMILY OF FULLERENES?

Alexander V. Bulgakov [a], Igor Ozerov [b], Wladimir Marine [b]

[a] Institute of Thermophysics, Prospect Lavrentyev 1, 630090 Novosibirsk, Russia,

e-mail: bulgakov@itp.nsc.ru

[b] Université de la Méditerranée, GPEC, UMR CNRS 6631, Marseille, France

ABSTRACT

Positively charged zinc oxide clusters $Zn_nO_m$ (up to n = 16, m ≤ n) of various stoichiometry were synthesized in the gas phase by excimer ArF laser ablation of a ZnO target and investigated using time-of-flight mass spectrometry. Depending on ablation conditions, either metal rich or stoichiometric clusters dominate in the mass spectrum. When the irradiated target surface is fairly fresh, the most abundant clusters are metal rich with $Zn_{n+1}O_n$ and $Zn_{n+3}O_n$ being the major series. The stoichiometric clusters are observed with an etched ablated surface. The magic numbers at n = 9, 11 and 15 in mass spectra of $(ZnO)_n$ clusters indicate that the clusters have hollow spheroid structures related to fullerenes. A local abundance minimum at n = 13 provides an additional evidence for the presence in the ablation plume of fullerene-like $(ZnO)_n$ clusters.

Recently much interest has been shown in ZnO nanostructured materials owing to their promising applications in chemical sensors, photocatalysis, nanoscale optoelectronic devices. A variety of ZnO nanocluster materials have been recently synthesized including nanorods [1] and nanotubes [2]. Bulk ZnO is an amphoteric oxide which exhibits both acid and basic properties. One might thus expect various structural forms of zinc oxide clusters with predominantly covalent or ionic chemical bonds depending on cluster size and stoichometry. It was theoretically predicted that $(ZnO)_n$ clusters with spheroid fullerene-like structures could be stable at n = 11, 12, and 15 [3]. Other calculations suggest stable spheroid structures for n = 8 and 9 [4]. However, the theoretically suggested structures have not yet been experimentally confirmed.

We report here on the observation of cationic $Zn_nO_m$ clusters of different stoichiometry in a laser-ablation plasma plume. The effect of ablation conditions on the relative cluster abundance is analyzed. The possible structures of clusters are discussed.

The apparatus used for laser ablation and cluster detection has been described previously [5]. In brief, a sintered ZnO disk was ablated under high vacuum conditions by a 15 ns ArF laser pulse at 193 nm. The laser fluence on the target was varied in the range 0.2 – 2 J/cm$^2$ at a fixed spot area of about 0.3 mm$^2$. At a distance of 57 mm from the target, positively charged particles of the laser-ablation plume were sampled parallel to the plume axis by a 500 V repeller pulse after a time delay, $t_d$, following the laser pulse. The relative abundance of the cations was analyzed using a reflectron time-of-flight (TOF) mass spectrometer. We distinguished roughly two different regimes in respect to cluster formation: (i) ablation of a relatively fresh surface (less than ~50 laser shots accumulated at the same spot), and (ii) ablation of an etched surface irradiated previously with at least 500 pulses. In all cases, we avoided considerable cratering at the surface.

At a relatively narrow range of laser fluence of $0.5 - 1$ J/cm$^2$ for both fresh and etched surfaces, a wide spectrum of zinc oxide clusters of various stoichiometry was observed. Figure 1a shows a mass spectrum of cationic species produced with a fresh ZnO surface at a laser fluence of 0.7 J/cm$^2$ corresponding to the maximum yield of clusters. The main $Zn_nO_m$ clusters are metal rich under these conditions and may be gathered into few series depending on the metal/oxygen ratio. The major series corresponds to $Zn_{n+1}O_n$ clusters with rather smooth size distribution. Clusters containing more than one excess metal atom are also present in high abundance and the $Zn_{n+3}O_n$ series is generally more abundant than $Zn_{n+2}O_n$ clusters. The stoichiometric clusters $(ZnO)_n$ with n = 2, 3, and 6-9 are present in low abundance under these conditions. The observation of metal-rich $Zn_nO_m$ clusters consists with the theoretically predicted "metallization" of small ionic clusters [6]. It has been speculated that anion vacancies are occupied by electrons donated from the excess metal atoms that results in stabilization of oxide cluster ions.

When the etched surface is ablated, the mass spectrum is significantly modified (Fig. 1b). The abundance of the stoichiometric clusters increases for n > 6 and becomes comparable to that of the $Zn_{n+1}O_n$ series. The $(ZnO)_n$ clusters contain mainly an odd number of zinc atoms with the magic numbers at n = 9, 11, and 15. However, there is a prominent local minimum at n = 13. The clusters are found to have nearly equal expansion velocity (~ 400 m/s under these conditions) over a wide mass range that implies that the clusters are formed due to gas-phase aggregation in the plasma plume rather than by direct emission from the surface [7].

The observed magic numbers and the fact that the stoichiometric $(ZnO)_n$ clusters dominate at fairly large sizes (n > 6) indicate, according to calculations [3,4], that the clusters have hollow cage-like structures related to fullerenes. The cages are made up of triply coordinated Zn and O atoms arranged with 6 rhombuses and an arbitrary number of hexagons.

The plausible structures for the magic $(ZnO)_n$ clusters are shown in Fig. 1b. A local abundance minimum at $n = 13$ provides an additional evidence for the presence of fullerene-like $(ZnO)_n$ clusters in the plume. It is reasonable to suggest that such clusters are more stable, the more the smaller polygons are isolated from each other (similar to the isolated pentagon rule for carbon fullerenes [8]). In the considered case of $(ZnO)_n$ cages, the structures with none of the rhombus to share an edge become geometrically possible at $n > 11$, except for $n = 13$. Moreover, the spherical structures at $n = 13$ are geometrically forced to have an additional rhombus or a 8-membered ring [9]. Therefore, the formation of cage-like $(ZnO)_{13}$ clusters in the kinetic pathway of zinc oxide cluster growth appears to be unfavorable since the adjusted four-membered rings impose the increased steric strains.

It is a pleasure to thank N.M. Bulgakova, N.A. Bulgakova, and V.I. Kosyakov for useful discussions. This work was supported by the Russian Foundation for Basic Research (grant 02-03-32221) and by the International Science and Technology Center (project 2310).

FIGURE CAPTION

Fig. 1. TOF mass spectra of cationic zinc oxide clusters produced by laser ablation of (a) fresh ZnO sample at fluence $E = 0.7$ J/cm$^2$ and time delay $t_d = 42$ μs and (b) etched ZnO surface at $E = 0.7$ J/cm$^2$ and $t_d = 42$ μs. The time delays correspond to maximum yields of clusters. The numbers above the peaks correspond to the numbers of Zn and O atoms, respectively. The insets show the mass spectra of low mass particles. Several peaks due to sample impurities are observed with the fresh sample. Plausible cage-like structures for the magic (ZnO)$_n$ clusters are shown in (b).

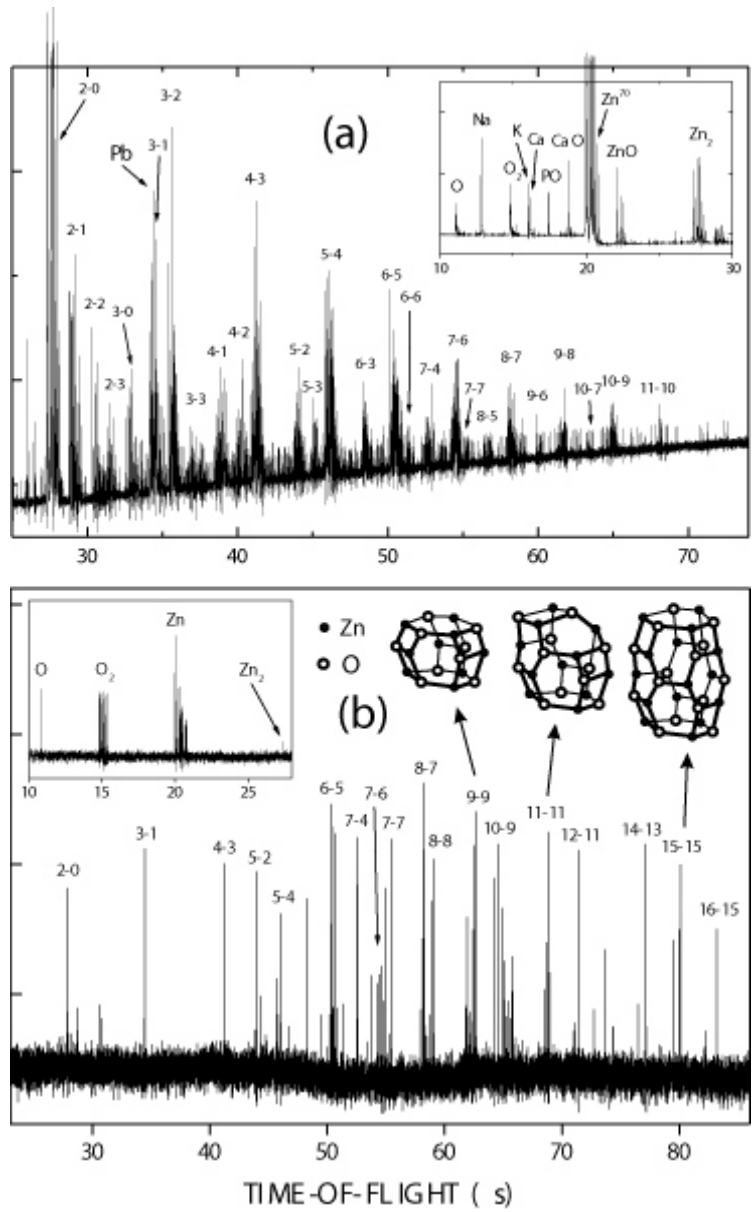

Fig 1